\documentclass[aps,pra,10pt,twocolumn,superscriptaddress,nofootinbib]{revtex4-2}
\ifdefined\pdfminorversion\pdfminorversion=7\fi
\usepackage{amsmath,amssymb,amsfonts,amsthm}
\usepackage{bm}
\usepackage{tikz}
\usepackage{graphicx}
\usepackage{booktabs}
\usetikzlibrary{arrows.meta}
\usepackage[colorlinks=true,linkcolor=blue,citecolor=blue,urlcolor=blue,hypertexnames=false]{hyperref}
\newcommand{\calH}{\mathcal{H}}
\newcommand{\calD}{\mathcal{D}}
\newcommand{\calF}{\mathcal{F}}
\newcommand{\calI}{\mathcal{I}}
\newcommand{\Tr}{\operatorname{Tr}}
\newcommand{\MA}{M_A}
\newcommand{\MAcr}{M_A^{\mathrm{cr}}}
\newcommand{\CA}{C_A}

\newtheorem{theorem}{Theorem}
\newtheorem{proposition}{Proposition}

\begin{document}
\raggedbottom \title{A Convex-Roof Affinity-Square Measure of Quantum Imaginarity} 
\author{Nan-Nan Guan} 
\affiliation{School of Science, East China University of Technology, Nanchang 330006, China} 
\author{Meng-Li Guo} 
\email{guomengli2020@163.com} 
\affiliation{School of Science, East China University of Technology, Nanchang 330006, China} 
\author{Xue-Na Zhu} 
\email{jing$_$feng1986@126.com} 
\affiliation{School of Mathematics and Statistics Science, Ludong University, Yantai 264025, China} 
\author{Zhi-Xiang Jin} 
\email{jzxjinzhixiang@126.com} 
\affiliation{School of Computer Science and Technology, Dongguan University of Technology, Dongguan, 523808, China} 
\author{Yan-Hui Zhou} 
\affiliation{Quantum Information Research Center and Jiangxi Province Key Laboratory of Applied Optical Technology, Shangrao Normal University, Shangrao 334001, China}

\begin{abstract}
Quantum imaginarity characterizes the part of a quantum state that cannot be represented by real density matrices in a fixed reference basis. We study an affinity-square measure of imaginarity for pure states and its convex-roof extension to mixed states. The pure state quantity is defined by minimizing the affinity-square measure over the convex set of real states. This optimization admits a closed expression determined by the modulus of the conjugation overlap between a state and its complex conjugate. The measure vanishes precisely when the state is real up to a global phase. We prove that the convex-roof extension is faithful, convex, monotone under real operations, and nonincreasing on average under selective real operations. For arbitrary single-qubit mixed states, we evaluate the convex roof exactly and express it in terms of the imaginary component of the off-diagonal density-matrix element. We also compare pure-state imaginarity with an auxiliary affinity-square coherence functional and give the corresponding phase weights.
\end{abstract}

\maketitle
\section{Introduction}
Quantum coherence is the resource associated with superposition in a fixed reference basis. The resource theory of coherence provides a framework for characterizing such basis-dependent nonclassicality and for quantifying its conversion under restricted operations \cite{Baumgratz,Streltsov2017}. Several coherence quantifiers and operational interpretations have been developed, including observable, geometric, operational, and robustness-based approaches \cite{Girolami,Pires,Winter2016,ctmm}. Coherence also plays a role in quantum thermodynamics and time-translation asymmetry \cite{y2,y5}. These results characterize the presence of off-diagonal components in a prescribed basis, but they do not by themselves separate real superpositions from superpositions that require intrinsically complex amplitudes. Connections between coherence and imaginarity have been studied in Refs.~\cite{Li2023RobustnessCoherence,Xu2025Coherence}.

This distinction motivates the resource theory of imaginarity. Standard quantum theory uses complex Hilbert spaces, whereas real-vector-space quantum theory imposes a structural restriction on states, correlations, and entanglement \cite{Wootters37,Hardy38,Aleksandrova39}. Network tests have shown that real-valued quantum theory can be ruled out in suitable experimental scenarios \cite{Renou2,Chen3,Li4,DianWu2022}. Further developments show that real and complex quantum theories can be discriminated efficiently and that complex amplitudes can yield operational advantages in communication tasks \cite{Batle2025Efficient,Elliott2025Advantage}. These results support a resource-theoretic treatment of the complex structure of quantum states. In this theory, the free states are density matrices that are real in the reference basis, and the free operations are quantum channels that admit real Kraus representations \cite{Hickey1,Wu1,Wu17}. Quantification, state conversion, distributed scenarios, Gaussian states, channel resources, broadcasting, erasure tasks, and witness-based detection have been studied within this framework \cite{Miyazaki,Gao,Xue,XuG,KangDa,Guo2024,Wu20261,Wu20262,Kondra2023RealOps,Zhang2023ChannelPower,Zhang2024Broadcasting,Shi2025Erasing,Sun2025Entanglement,Fernandes2024Witness}.

A geometric resource measure compares a state with the corresponding free-state set \cite{Chitambar2019}. For imaginarity, the relevant free set is the convex set of real density matrices. A geometric construction should therefore quantify the deviation of a state from this real sector. Hellinger-distance-based coherence and related affinity-type quantities provide useful tools for such comparisons \cite{Jin,Luo2004,Pitrik2020,Orthey2025}. Entropy-based and distance-induced measures of imaginarity have also been proposed in recent studies \cite{Xu2024Tsallis,Guo2025Distance,Du2025PureState,Tian2025JSD}. However, a distance-like expression does not automatically satisfy the axioms of a resource monotone. A valid measure of imaginarity must vanish exactly on real states, decrease under real operations, and remain nonincreasing on average under selective real operations.

In this work, we study a convex-roof affinity-square measure of quantum imaginarity. The pure state quantity is defined by minimizing the squared-affinity deviation from the full real-state set. The optimum has a closed form in terms of the modulus of the conjugation overlap. The convex-roof extension assigns to a mixed state the minimum average pure state imaginarity over all pure state decompositions. This choice of pure-state function falls within the convex-roof construction of Du and Bai~\cite{Du2025PureState}. We prove that this extension is faithful, convex, monotone under real operations, and nonincreasing on average under selective real operations. We further analyze its behavior under coherent superposition and show that the relevant interference term is the bilinear overlap induced by complex conjugation. Comparison with an auxiliary affinity-square coherence functional gives a pure-state relation expressed through pairwise phase differences. Finally, we evaluate the convex roof exactly for arbitrary single-qubit mixed states, where the measure depends only on the imaginary component of the off-diagonal density-matrix element and is related explicitly to robustness, the $l_1$-norm imaginarity, trace-distance imaginarity, and relative entropy of imaginarity.

\section{Affinity-square convex-roof imaginarity}
\label{sec:affinity}
Let ${|j\rangle}_{j=0}^{d-1}$ be a fixed orthonormal basis of a finite-dimensional Hilbert space $\calH$, and let $\calD(\calH)$ denote the set of density operators on $\calH$. In the resource theory of imaginarity, the free states are the density matrices whose entries are real in this basis \cite{Hickey1,Wu1,Wu17}. Thus
\begin{equation*}
\calF=
\left\{
\rho\in\calD(\calH):
\langle m|\rho|n\rangle\in\mathbb{R},\ \forall m,n
\right\}.
\label{eq:free-states}
\end{equation*}
Equivalently, $\rho\in\calF$ if and only if $\rho=\rho^*$ in the reference basis. Hence each free state is a real symmetric positive semidefinite matrix with unit trace. The free operations are real operations. These are completely positive trace-preserving maps that admit at least one Kraus representation $\varepsilon(\rho)=\sum_jK_j\rho K_j^\dagger$, where every Kraus matrix $K_j$ is real in the same reference basis. Such operations preserve $\calF$: if $\sigma\in\calF$, then each matrix $K_j\sigma K_j^\dagger$ is real, and therefore $\varepsilon(\sigma)\in\calF$. A valid measure of imaginarity should satisfy the standard resource-theoretic requirements \cite{Wu17,Hickey1}. \textit{Faithfulness}: $\mathcal{M}(\rho) \geq 0$, with equality if and only if $\rho \in \mathcal{F}$; \textit{Monotonicity under free operations}: $\mathcal{M}(\varepsilon(\rho)) \leq \mathcal{M}(\rho)$ for any real operation $\varepsilon$; \textit{Nonincrease on average under selective free operations}: $\sum_j p_j \mathcal{M}(\rho_j) \leq \mathcal{M}(\rho)$, where $p_j = \mathrm{Tr}(K_j \rho K_j^\dagger)$, $\rho_j = K_j \rho K_j^\dagger / p_j$, and ${K_j}$ are real Kraus operators; \textit{Convexity}: $\mathcal{M}\big(\sum_j p_j \rho_j\big) \leq \sum_j p_j \mathcal{M}(\rho_j)$ for any probability distribution ${p_j}$ and states ${\rho_j}$.

We next define the geometric quantity used to compare a state with the real-state set $\calF$. For density operators $\rho,\sigma\in\calD(\calH)$, the quantum affinity is
\begin{equation*}
A(\rho,\sigma)=\Tr(\sqrt{\rho}\sqrt{\sigma}),
\label{eq:affinity}
\end{equation*}
It satisfies $0\leq A(\rho,\sigma)\leq1$, with equality $A(\rho,\sigma)=1$ if and only if $\rho=\sigma$. We use the associated affinity-square measure
\begin{equation*}
d_A(\rho,\sigma)=1-A(\rho,\sigma)^2 .
\label{eq:affinity-square-distance}
\end{equation*}
In this work, $d_A$ is used as a resource-theoretic geometric functional, not as a metric. The arguments below require only its positivity, normalization, and the fact that its pure state optimization over $\calF$ can be evaluated in closed form. This closed form provides the pure state input for the convex-roof construction.

For a pure state $|\psi\rangle$, we define the affinity-square imaginarity as
\begin{equation}
\MA(|\psi\rangle)=\min_{\sigma\in\calF}
d_A(|\psi\rangle\langle\psi|,\sigma).
\label{eq:pure-MA-def}
\end{equation}
The minimization is taken over all real density matrices, not only over real pure states. Hence $\MA(|\psi\rangle)$ gives the minimal affinity-square measure from the real-state set. Since the definition depends on the projector $|\psi\rangle\langle\psi|$, the value of $\MA(|\psi\rangle)$ is invariant under a global phase of $|\psi\rangle$. At the same time, it remains sensitive to non-real relative phases in the reference basis.

\begin{figure}[!t]
	\centering
	\includegraphics[width=0.78\columnwidth]{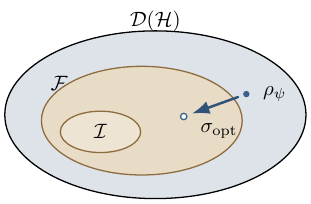}
\caption{Geometry of the affinity-square construction. The pure state optimization is performed over the full set $\calF$ of real states, which contains the incoherent set $\calI$. The convex roof extends the resulting pure state quantity to mixed states by minimizing the average over all pure state decompositions.}
	\label{fig:affinity-square-convex-roof}
\end{figure}

For a mixed state $\rho$, we extend the pure state quantity by the convex roof,
\begin{equation}
\MAcr(\rho)
=
\inf_{\{p_k,|\psi_k\rangle\}}
\sum_k p_k \MA(|\psi_k\rangle),
\label{eq:convex-roof-def}
\end{equation}
where the infimum is over all pure state decompositions $\rho=\sum_kp_k|\psi_k\rangle\langle\psi_k|$. Thus $\MAcr(\rho)$ is the minimum average pure state imaginarity among all ensembles that realize the same density operator. By construction, this definition gives a convex mixed state extension of the pure state measure. Figure~\ref{fig:affinity-square-convex-roof} summarizes the geometry. For a pure input, the affinity-square measure is minimized over the full real-state set $\calF$. The incoherent states form a subset of $\calF$, which also contains real coherent states. In finite dimensions, the infimum in Eq.~\eqref{eq:convex-roof-def} can be taken over finite ensembles. Indeed, the real vector space of Hermitian trace-one operators on a $d$-dimensional Hilbert space has dimension $d^2-1$. By Carath\'eodory's theorem, any density operator can be represented as a convex combination of at most $d^2$ pure state projectors \cite{Rockafellar1970ConvexAnalysis}. Hence the convex-roof optimization may be restricted to ensembles with at most $d^2$ pure states. After padding with zero probabilities, the feasible ensembles form a closed subset of the compact space $\Delta_{d^2}\times (\mathbb{C}P^{d-1})^{d^2}.$ Since $\MA(|\psi\rangle)$ is continuous on the pure state manifold, the objective $\sum_k p_k\MA(|\psi_k\rangle)$ is continuous on this feasible set. Therefore the infimum is attained.

The following proposition gives the pure state formula that serves as the building block of the convex roof. It expresses the measure in terms of the overlap between a state and its complex conjugate and therefore isolates the non-real phase contribution.
\begin{proposition}
\label{prop:pure-formula}
For every pure state $|\psi\rangle\in\calH$,
\begin{equation}
\MA(|\psi\rangle)=\frac{1-|\langle\psi^*|\psi\rangle|^2}{2},
\label{eq:pure-MA}
\end{equation}
where $|\psi^*\rangle$ denotes the complex conjugate of $|\psi\rangle$ in the fixed reference basis.
\end{proposition}

{\sf [Proof]} The minimization in Eq.~\eqref{eq:pure-MA-def} is taken over the full convex set of real density matrices. Set $r=|\langle\psi^*|\psi\rangle|.$ If $d=1$, every normalized pure state is real up to a global phase. Hence $r=1$, the state is free, and the claimed formula gives $M_A(|\psi\rangle)=0$.

Assume henceforth that $d\ge 2$. Since the projector $|\psi\rangle\langle\psi|$ is invariant under a global phase, we may choose the phase of $|\psi\rangle$ such that $\langle\psi^*|\psi\rangle=\psi^T\psi=r\geq0 .$ Write $|\psi\rangle=x+iy,$ where $x$ and $y$ are real vectors in the fixed reference basis. The normalization of $|\psi\rangle$ and the phase choice above give $x^Ty=0$, $\|x\|^2+\|y\|^2=1$, $\|x\|^2-\|y\|^2=r$. Hence
\begin{equation*}
\|x\|^2=\frac{1+r}{2},
\qquad
\|y\|^2=\frac{1-r}{2}.
\end{equation*}
Let $O$ be a real orthogonal transformation that maps the two orthogonal real vectors $x$ and $y$ into the span of $|0\rangle$ and $|1\rangle$. Real orthogonal conjugation preserves the real-state set: $O\mathcal F O^T=\mathcal F.$ Moreover, the affinity is invariant under unitary conjugation: $A\!\left(O|\psi\rangle\langle\psi|O^T,O\sigma O^T\right)=A\!\left(|\psi\rangle\langle\psi|,\sigma\right).$ Therefore the optimization value is invariant under such transformations. The problem is thus equivalent to the one for
\begin{equation*}
|u\rangle
=
a|0\rangle+ib|1\rangle,
\qquad
a^2=\frac{1+r}{2},
~b^2=\frac{1-r}{2}.
\end{equation*}
The degenerate cases are included in this expression by taking $a=0$ or $b=0$ when appropriate.

Let $\sigma\in\calF$ and set $T=\sqrt{\sigma}.$ Since $\sigma$ is real symmetric and positive semidefinite, $T$ is also real symmetric and positive semidefinite. Moreover, $\Tr T^2=\Tr\sigma=1.$ Conversely, every real positive semidefinite matrix $T$ with $\Tr T^2=1$ defines an admissible real state $\sigma=T^2$.

For $|u\rangle=a|0\rangle+ib|1\rangle$, the affinity is
\begin{equation*}
A(|u\rangle\langle u|,\sigma)
=
\langle u|T|u\rangle
=
a^2T_{00}+b^2T_{11},
\end{equation*}
because the off-diagonal terms cancel for real symmetric $T$. By the Cauchy--Schwarz inequality,
\begin{align*}
a^2T_{00}+b^2T_{11}
&\leq
\sqrt{a^4+b^4}\,
\sqrt{T_{00}^2+T_{11}^2}
\\
&\leq
\sqrt{a^4+b^4},
\end{align*}
where the second inequality follows from $T_{00}^2+T_{11}^2\leq \Tr T^2=1$. The bound is attained by
\begin{equation*}
T_*
=
\frac{
	a^2|0\rangle\langle0|
	+
	b^2|1\rangle\langle1|
}
{\sqrt{a^4+b^4}} .
\end{equation*}
Indeed, $T_*$ is real positive semidefinite and satisfies $\Tr T_*^2=1.$ Thus $\sigma_*=T_*^2$ is a real density matrix, and
\begin{equation*}
A(|u\rangle\langle u|,\sigma_*)
=
\sqrt{a^4+b^4}.
\end{equation*}
Therefore
\begin{equation*}
\max_{\sigma\in\calF}
A(|u\rangle\langle u|,\sigma)
=
\sqrt{a^4+b^4}.
\end{equation*}

Using Eq.~\eqref{eq:pure-MA-def}, we obtain
\begin{align*}
\MA(|\psi\rangle)
&=
1-
\max_{\sigma\in\calF}
A(|\psi\rangle\langle\psi|,\sigma)^2
\\
&=
1-(a^4+b^4).
\end{align*}
Since
\begin{equation*}
a^4+b^4
=
\left(\frac{1+r}{2}\right)^2
+
\left(\frac{1-r}{2}\right)^2
=
\frac{1+r^2}{2},
\end{equation*}
we have
\begin{equation*}
\MA(|\psi\rangle)
=
\frac{1-r^2}{2}
=
\frac{1-|\langle\psi^*|\psi\rangle|^2}{2}.
\end{equation*}
This proves Eq.~\eqref{eq:pure-MA}. \hfill $\square$

Equation~\eqref{eq:pure-MA} specifies the phase information detected by $\MA$. The complex bilinear overlap $\langle\psi^*|\psi\rangle=\psi^T\psi$ compares a pure state with its complex conjugate in the fixed basis. Under a global phase transformation $|\psi\rangle\mapsto e^{i\alpha}|\psi\rangle$, this overlap transforms as $\langle\psi^*|\psi\rangle \mapsto e^{2i\alpha}\langle\psi^*|\psi\rangle,$ whereas its modulus is invariant. Hence $\MA(|\psi\rangle)=0$ if and only if $|\psi\rangle$ can be made real by a global phase, namely if and only if $|\psi\rangle$ is a free pure state. A nonzero value therefore detects relative phases that cannot be removed by a global phase and cannot be represented within real quantum theory. Thus $\MA$ does not measure basis superposition itself; it measures the non-real part of such a superposition. For $d\geq2$, the maximal pure state value is $\frac{1}{2}$, attained when $\langle\psi^*|\psi\rangle=0$, for example by $\frac{1}{\sqrt2}(|0\rangle+i|1\rangle)$.

The closed form in Eq.~\eqref{eq:pure-MA} reduces the pure state analysis to the behavior of the bilinear conjugation overlap under real Kraus operators. This reduction provides the starting point for the resource-theoretic proof of the convex-roof extension. We first prove selective monotonicity for pure inputs and then use the convex-roof construction to extend the result to arbitrary mixed states.
\begin{theorem}
\label{thm:valid-measure}
The convex-roof affinity-square quantity
\begin{equation*}
\MAcr(\rho)
=
\inf_{\{p_k,|\psi_k\rangle\}}
\sum_k p_k \MA(|\psi_k\rangle)
\label{eq:MAcr-theorem}
\end{equation*}
is a valid measure of imaginarity on $\calD(\calH)$.
\end{theorem}

{\sf [Proof]} We first prove selective monotonicity for pure input states. Let $|\psi\rangle$ be a pure state and let $\{K_j\}$ be a real Kraus representation of a trace-preserving real operation. For each branch with $p_j>0$, set
\begin{equation*}
p_j
=
\Tr(K_j|\psi\rangle\langle\psi|K_j^\dagger),
\qquad
|\psi_j\rangle
=
\frac{K_j|\psi\rangle}{\sqrt{p_j}} .
\end{equation*}
Branches with $p_j=0$ are omitted. Define the bilinear conjugation overlap $S(\psi)=\langle\psi^*|\psi\rangle=\psi^T\psi,$ $\alpha_j=\psi^TK_j^TK_j\psi .$ Since each $K_j$ is real, $K_j^\dagger=K_j^T$, and therefore $S(\psi_j)=\psi_j^T\psi_j=\frac{\alpha_j}{p_j}.$ The trace-preserving condition gives $\sum_jK_j^TK_j=I$, and $\sum_j\alpha_j=S(\psi)$. Using Proposition~\ref{prop:pure-formula}, we obtain
\begin{align*}
\sum_jp_j\MA(|\psi_j\rangle)
&=
\frac12\sum_jp_j
\Big(
1-\left|S(\psi_j)\right|^2
\Big)
\nonumber\\
&=
\frac12\sum_jp_j
\Big(
1-\frac{|\alpha_j|^2}{p_j^2}
\Big)
\nonumber\\
&=
\frac12
\Big(
1-\sum_j\frac{|\alpha_j|^2}{p_j}
\Big).
\end{align*}
The Cauchy--Schwarz inequality gives
\begin{align*}
|S(\psi)|^2
&=
\Big|
\sum_j\alpha_j
\Big|^2=
\Big|
\sum_j
\sqrt{p_j}\,
\frac{\alpha_j}{\sqrt{p_j}}
\Big|^2
\nonumber\\
&\leq
\Big(
\sum_jp_j
\Big)
\Big(
\sum_j\frac{|\alpha_j|^2}{p_j}
\Big)=
\sum_j\frac{|\alpha_j|^2}{p_j}.
\end{align*}
Substitution of this bound into the preceding expression yields
\begin{equation}
\sum_jp_j\MA(|\psi_j\rangle)
\leq
\frac12\left(1-|S(\psi)|^2\right)
=
\MA(|\psi\rangle).
\label{eq:pure-selective-monotonicity}
\end{equation}
Thus $\MA$ is nonincreasing on average under selective real operations for pure states.

We now extend this result to mixed states by the convex roof.  Since the convex-roof infimum is attained in finite dimensions, choose an optimal pure state decomposition $\rho=\sum_k r_k|\psi_k\rangle\langle\psi_k|$ such that
\begin{equation*}
\sum_k r_k\MA(|\psi_k\rangle)
=
\MAcr(\rho).
\end{equation*}
For the same selective real operation, define
\begin{equation*}
p_{j|k}
=
\Tr(K_j|\psi_k\rangle\langle\psi_k|K_j^\dagger),
\qquad
|\psi_{j|k}\rangle
=
\frac{K_j|\psi_k\rangle}{\sqrt{p_{j|k}}},
\end{equation*}
whenever $p_{j|k}>0$. The probability of branch $j$ is $p_j=\sum_k r_kp_{j|k},$ and, for $p_j>0$, the normalized output state is
\begin{equation*}
\rho_j
=
\frac{K_j\rho K_j^\dagger}{p_j}
=
\sum_k
\frac{r_kp_{j|k}}{p_j}
|\psi_{j|k}\rangle\langle\psi_{j|k}|.
\end{equation*}
Thus the displayed ensemble is an admissible pure-state decomposition of $\rho_j$. By the definition of the convex roof,
\begin{equation*}
\MAcr(\rho_j)
\leq
\sum_k
\frac{r_kp_{j|k}}{p_j}
\MA(|\psi_{j|k}\rangle).
\end{equation*}
Multiplying by $p_j$ and summing over all nonzero branches gives
\begin{align*}
\sum_jp_j\MAcr(\rho_j)
&\leq
\sum_{j,k}
r_kp_{j|k}\MA(|\psi_{j|k}\rangle)
\\
&=
\sum_k r_k
\sum_jp_{j|k}\MA(|\psi_{j|k}\rangle)
\\
&\leq
\sum_k r_k\MA(|\psi_k\rangle)=\MAcr(\rho).
\end{align*}
The third line follows from the pure-state selective monotonicity in Eq.~\eqref{eq:pure-selective-monotonicity}. Hence
\begin{equation}
\sum_jp_j\MAcr(\rho_j)
\leq
\MAcr(\rho).
\label{eq:mixed-selective-monotonicity}
\end{equation}
Therefore $\MAcr$ is nonincreasing on average under selective real operations.

Next, we prove convexity. Let $\rho=\sum_l q_l\rho_l$ with $q_l\geq0$ and $\sum_lq_l=1$. For each $\rho_l$, choose an optimal pure-state decomposition $\rho_l=\sum_k p_{k|l} |\psi_{k|l}\rangle\langle\psi_{k|l}|$ satisfying
\begin{equation*}
\sum_kp_{k|l}\MA(|\psi_{k|l}\rangle)
=
\MAcr(\rho_l).
\end{equation*}
Then $\{q_lp_{k|l},|\psi_{k|l}\rangle\}_{l,k}$ is a pure-state decomposition of $\rho$. Therefore
\begin{align*}
\MAcr(\rho)
&\leq
\sum_{l,k}q_lp_{k|l}\MA(|\psi_{k|l}\rangle)=
\sum_lq_l\MAcr(\rho_l).
\end{align*}
Thus
\begin{equation}
\MAcr\!\Big(\sum_lq_l\rho_l\Big)
\leq
\sum_lq_l\MAcr(\rho_l).
\label{eq:convexity}
\end{equation}

Nonselective monotonicity follows from convexity and selective monotonicity. If $\varepsilon(\rho)=\sum_jK_j\rho K_j^\dagger=\sum_jp_j\rho_j$ is the output obtained after discarding the branch label, then
\begin{equation}
\MAcr[\varepsilon(\rho)]
\leq
\sum_jp_j\MAcr(\rho_j)
\leq
\MAcr(\rho).
\label{eq:nonselective-monotonicity}
\end{equation}

It remains to prove faithfulness. If $\rho\in\calF$, then $\rho$ is a real symmetric positive semidefinite matrix with unit trace. Hence it admits a spectral decomposition $\rho=\sum_k\lambda_k|e_k\rangle\langle e_k|$ with real orthonormal eigenvectors $|e_k\rangle$. Each $|e_k\rangle$ is a free pure state, so $\MA(|e_k\rangle)=0$. Therefore $\MAcr(\rho)=0.$ Conversely, suppose that $\MAcr(\rho)=0$. Since the convex-roof minimum is attained, there exists a pure-state decomposition $\rho=\sum_kp_k|\psi_k\rangle\langle\psi_k|$ such that $\sum_kp_k\MA(|\psi_k\rangle)=0.$ Each summand is non-negative. Hence every component with $p_k>0$ satisfies $\MA(|\psi_k\rangle)=0.$ By Proposition~\ref{prop:pure-formula}, $|\langle\psi_k^*|\psi_k\rangle|=1.$ Write the components of $|\psi_k\rangle$ as $(\psi_k)_j=|(\psi_k)_j|e^{i\theta_j}.$ Then $\Big|\sum_j |(\psi_k)_j|^2 e^{2i\theta_j}\Big|=1.$ The expression inside the modulus is a convex combination of unit complex numbers. Equality in the triangle inequality holds only if all nonzero terms have the same phase. Hence all nonzero amplitudes of $|\psi_k\rangle$ have a common phase modulo $\pi$. Thus each $|\psi_k\rangle$ is real up to a global phase, and each projector $|\psi_k\rangle\langle\psi_k|$ belongs to $\calF$. Since $\calF$ is convex, $\rho\in\calF.$ We have therefore shown that
\begin{equation}
\MAcr(\rho)=0
\quad\Longleftrightarrow\quad
\rho\in\calF .
\label{eq:faithfulness}
\end{equation}

Equations~\eqref{eq:mixed-selective-monotonicity}, \eqref{eq:convexity}, \eqref{eq:nonselective-monotonicity}, and \eqref{eq:faithfulness} establish selective monotonicity, convexity, nonselective monotonicity, and faithfulness, respectively. Thus $\MAcr$ is a valid measure of imaginarity. \hfill $\square$

Theorem~\ref{thm:valid-measure} establishes that $\MAcr$ satisfies the standard resource-theoretic conditions. The pure state formula in Proposition~\ref{prop:pure-formula} also identifies the quantity that controls the resource. It depends on the complex bilinear form $\psi^T\psi$, not only on the Hermitian normalization $\langle\psi|\psi\rangle$. This distinction matters for coherent superpositions, because interference in the conjugation overlap can change the imaginarity while the component states remain fixed. The next result makes this dependence explicit.

\begin{theorem}\label{thm:superposition}
Let $|\phi_1\rangle$ and $|\phi_2\rangle$ be normalized pure states satisfying $\langle\phi_1|\phi_2\rangle=0$. Consider the normalized superposition $|\phi\rangle=c_1|\phi_1\rangle+c_2|\phi_2\rangle$ with $|c_1|^2+|c_2|^2=1$. All complex conjugations are taken in the fixed reference basis. Define $s_i=\langle\phi_i^*|\phi_i\rangle$, $r_i=|s_i|$, $i=1,2$, and $t=\langle\phi_1^*|\phi_2\rangle=\langle\phi_2^*|\phi_1\rangle$. Then
\begin{equation}
\MA(|\phi\rangle)
=
\frac{1-|S|^2}{2},
\label{eq:superposition-exact}
\end{equation}
where $S=c_1^2s_1+c_2^2s_2+2c_1c_2t .$ Moreover, let
\begin{equation}
U=
\min
\left\{
1,\,
|c_1|^2r_1+|c_2|^2r_2+2|c_1c_2||t|
\right\}\nonumber
\end{equation}
and
\begin{equation}
L=
\max
\left\{
0,\,
\bigl||c_1|^2r_1-|c_2|^2r_2\bigr|
-2|c_1c_2||t|
\right\}.
\nonumber
\end{equation}
Then
\begin{equation}
\frac{1-U^2}{2}
\leq
\MA(|\phi\rangle)
\leq
\frac{1-L^2}{2}.
\label{eq:superposition-bound}
\end{equation}	
\end{theorem}

{\sf [Proof]} The conjugation overlap of the superposition is computed with the complex bilinear form:
\begin{align}
\langle\phi^*|\phi\rangle
&=
(c_1\phi_1+c_2\phi_2)^T
(c_1\phi_1+c_2\phi_2)
\nonumber\\
&=
c_1^2\phi_1^T\phi_1
+
c_2^2\phi_2^T\phi_2
+
2c_1c_2\phi_1^T\phi_2
\nonumber\\
&=
c_1^2s_1+c_2^2s_2+2c_1c_2t .
\nonumber
\end{align}
Since $\phi_1^T\phi_2=\phi_2^T\phi_1$, the two cross terms are equal, which gives the stated definition of $t$ and $S$. Equation~\eqref{eq:superposition-exact} follows from Proposition~\ref{prop:pure-formula}. It remains to bound $|S|$. The triangle inequality gives
\begin{equation}
|S|\leq |c_1|^2r_1+|c_2|^2r_2+2|c_1c_2||t|.\nonumber
\end{equation}
Together with the general bound $|S|\leq1$, this gives $|S|\leq U$. The reverse triangle inequality applied to the three terms in $S$ gives
\begin{equation}
|S|\geq \bigl||c_1|^2r_1-|c_2|^2r_2\bigr|-2|c_1c_2||t|.\nonumber
\end{equation}
Combining this estimate with $|S|\geq0$ gives $L\leq |S|$. Hence $L\leq |S|\leq U .$ Since the function $(1-x^2)/2$ decreases on $0\leq x\leq1$, Eq.~\eqref{eq:superposition-bound} follows. \hfill $\square$

Theorem~\ref{thm:superposition} shows that the imaginarity of a superposed state is governed by the bilinear conjugation overlap $S$. Thus the relative phase between the two components can either suppress or enhance $\MA(|\phi\rangle)$ through the cross term $2c_1c_2t$. The following qutrit example illustrates this phase dependence and verifies the bounds in Eq.~\eqref{eq:superposition-bound}.

\begin{figure*}[!t]
	\centering
	\includegraphics[width=0.95\textwidth]{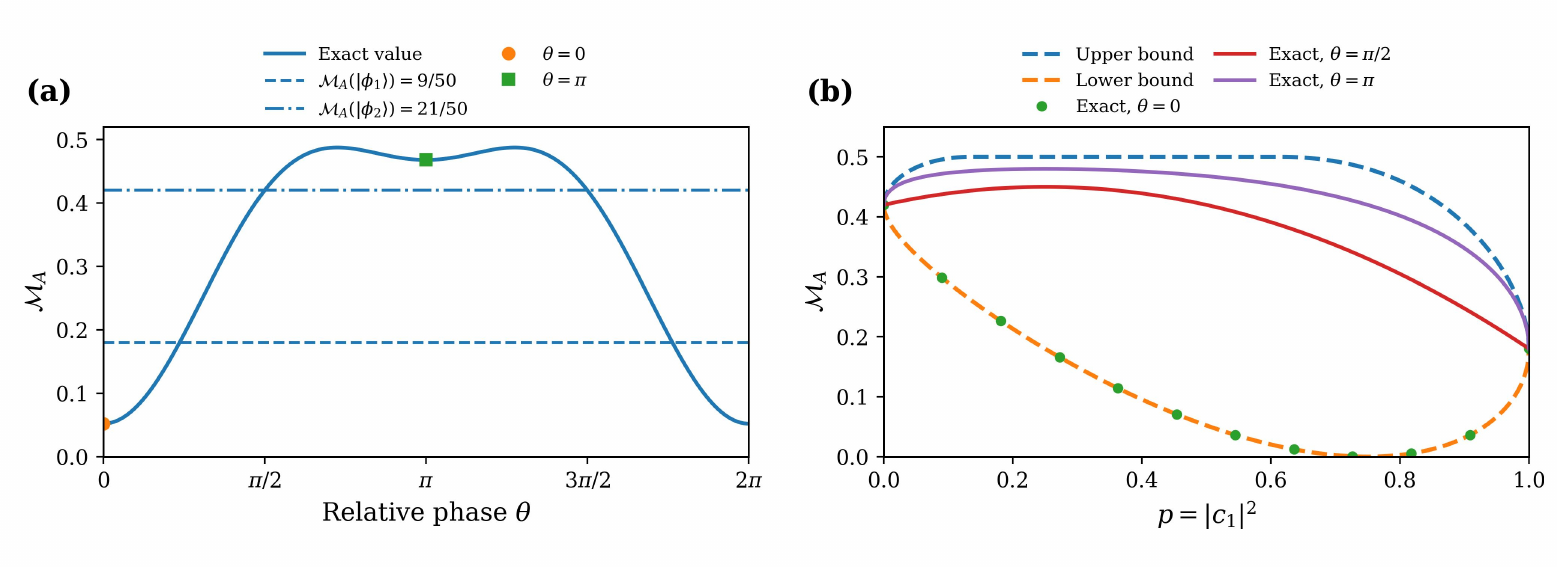}
\caption{Qutrit superposition example. (a) Exact value of $\MA(|\phi(1/2,\theta)\rangle)$ as a function of the relative phase $\theta$. The horizontal dashed and dash-dotted lines mark $\MA(|\phi_1\rangle)=9/50$ and $\MA(|\phi_2\rangle)=21/50$, respectively. The orange circle and green square mark $\theta=0$ and $\theta=\pi$. (b) Bounds in Eq.~\eqref{eq:superposition-bound} as a function of $p=|c_1|^2$. The red and purple solid curves show exact values for $\theta=\pi/2$ and $\theta=\pi$, respectively; the green points show $\theta=0$. The blue and orange dashed curves give the upper and lower bounds.}
	\label{fig:qutrit-superposition-example}
\end{figure*}

To illustrate the phase sensitivity of $\MA$ and the bounds in Theorem~\ref{thm:superposition}, consider a qutrit with reference basis $\{|0\rangle,|1\rangle,|2\rangle\}$. Define
\begin{align}
|\phi_1\rangle&=\frac{3}{\sqrt{10}}|0\rangle+\frac{i}{\sqrt{10}}|1\rangle+0|2\rangle,
\nonumber\\
|\phi_2\rangle&=\frac{1}{\sqrt{30}}|0\rangle-i\sqrt{\frac{3}{10}}|1\rangle+\sqrt{\frac{2}{3}}|2\rangle.\nonumber
\end{align}
Both states are normalized. Their Hermitian inner product is
\begin{equation}
\langle\phi_1|\phi_2\rangle
=
\frac{3}{\sqrt{10}}\frac{1}{\sqrt{30}}
+
\left(-\frac{i}{\sqrt{10}}\right)
\left(-i\sqrt{\frac{3}{10}}\right)
=
0 .
\nonumber
\end{equation}
so the assumptions of Theorem~\ref{thm:superposition} are satisfied. The corresponding conjugation overlaps are
\begin{equation}
s_1=\frac45,
\qquad
s_2=\frac25,
\qquad
t=\frac{\sqrt3}{5}.
\label{eq:qutrit-example-overlaps}
\end{equation}
Thus $r_1=\frac{4}{5}$, $r_2=\frac{2}{5}$, and $|t|=\frac{\sqrt3}{5}$. Proposition~\ref{prop:pure-formula} gives
\begin{equation*}
\MA(|\phi_1\rangle)
=
\frac{1-r_1^2}{2}
=
\frac{9}{50},
\qquad
\MA(|\phi_2\rangle)
=
\frac{1-r_2^2}{2}
=
\frac{21}{50}.
\end{equation*}
For $0\leq p\leq1$, introduce the normalized superposition
\begin{equation*}
|\phi(p,\theta)\rangle
=
\sqrt{p}\,|\phi_1\rangle
+
e^{i\theta}\sqrt{1-p}\,|\phi_2\rangle .
\end{equation*}
Its normalization follows from $\langle\phi_1|\phi_2\rangle=0$. Substituting Eq.~\eqref{eq:qutrit-example-overlaps} into $S$ yields
\begin{equation*}
S(p,\theta)
=
\frac{4p}{5}
+
\frac{2(1-p)}{5}e^{2i\theta}
+
\frac{2\sqrt3}{5}\sqrt{p(1-p)}\,e^{i\theta}.
\end{equation*}
Therefore,
\begin{equation*}
\MA(|\phi(p,\theta)\rangle)
=
\frac{1-|S(p,\theta)|^2}{2}.
\end{equation*}
For the balanced superposition $p=\frac{1}{2}$,
\begin{equation*}
S\Big(\frac12,\theta\Big)
=
\frac25
+
\frac15 e^{2i\theta}
+
\frac{\sqrt3}{5}e^{i\theta}.
\end{equation*}
A direct evaluation gives
\begin{equation*}
\MA\Big(|\phi(\frac{1}{2},\theta)\rangle\Big)
=
\frac{21-6\sqrt3\cos\theta-8\cos^2\theta}{50}.
\end{equation*}
This expression shows that the relative phase changes the imaginarity even when the two component states and their weights are fixed. In particular,
\begin{align*}
\MA\Big(|\phi(\frac{1}{2},0)\rangle\Big)
&=
\frac{13-6\sqrt3}{50}
\approx
0.0522,\nonumber\\
\MA\Big(|\phi(\frac{1}{2},\pi)\rangle\Big)
&=
\frac{13+6\sqrt3}{50}
\approx
0.4678 .
\end{align*}
The same pair of component states can therefore produce a small or nearly maximal qutrit value, depending only on the relative phase.

The bounds in Eq.~\eqref{eq:superposition-bound} take a simple form in this example. From Eq.~\eqref{eq:qutrit-example-overlaps},
\begin{equation*}
U(p)
=
\min
\left\{
1,\,
\frac{4p}{5}
+
\frac{2(1-p)}{5}
+
\frac{2\sqrt3}{5}\sqrt{p(1-p)}
\right\},
\end{equation*}
and
\begin{equation*}
L(p)
=
\max
\left\{
0,\,
\left|
\frac{4p}{5}
-
\frac{2(1-p)}{5}
\right|
-
\frac{2\sqrt3}{5}\sqrt{p(1-p)}
\right\}.
\end{equation*}
Consequently,
\begin{equation}
\frac{1-U(p)^2}{2}\leq
\MA(|\phi(p,\theta)\rangle)
\leq
\frac{1-L(p)^2}{2}.
\label{eq:qutrit-bound-verification}
\end{equation}
The bounds depend only on $p$, $r_1$, $r_2$, and $|t|$, whereas the exact value also depends on the relative phase $\theta$. Thus the example separates phase-independent amplitude constraints from phase-dependent interference in $S(p,\theta)$. Figure~\ref{fig:qutrit-superposition-example} displays the phase dependence for the balanced case and the corresponding bound verification as functions of $p$.

\section{Comparison with coherence and other imaginarity measures}
\label{sec:comparison}
The preceding sections establish $\MAcr$ as an imaginarity monotone and show that, for pure states, $\MA$ is determined by the conjugation overlap $\langle\psi^*|\psi\rangle$. We now compare this quantity with coherence and with standard measures of imaginarity. We first compare the two pure-state affinity-square functionals, then evaluate the qubit convex roof and relate it to existing quantifiers.

\subsection{Pure-state relation to coherence}
\label{sec:coherence-hierarchy}
The relation between coherence and imaginarity has been studied using, for example, robustness-based comparisons and the difference between a state and its real part \cite{Li2023RobustnessCoherence,Xu2025Coherence}. The pure state formula in Proposition~\ref{prop:pure-formula} gives a direct comparison within the same affinity-square geometry. Consider
\begin{equation}
|\psi\rangle
=
\sum_j\sqrt{p_j}e^{i\theta_j}|j\rangle,
\qquad
p_j\geq0,~
\sum_jp_j=1 .
\nonumber
\end{equation}
To compare imaginarity with basis coherence, define the auxiliary pure state coherence functional
\begin{equation}
\CA(|\psi\rangle)=\min_{\delta\in\calI}
\Big[1-A(|\psi\rangle\langle\psi|,\delta)^2\Big],
\nonumber
\end{equation}
where $\calI$ is the set of incoherent states, i.e., density matrices diagonal in the reference basis. This functional is obtained from Eq.~\eqref{eq:pure-MA-def} by replacing the real-state set $\calF$ with $\calI$. Here $\CA$ serves only as a pure-state reference functional, not as a general coherence monotone. We do not consider a mixed-state extension or its resource-theoretic properties. For an incoherent state $\delta=\sum_jq_j|j\rangle\langle j|$, with $q_j\geq0$ and $\sum_jq_j=1$, one has
\begin{equation}
A(|\psi\rangle\langle\psi|,\delta)
=
\sum_jp_j\sqrt{q_j}.
\nonumber
\end{equation}
The Cauchy--Schwarz inequality gives
\begin{equation}
\Big(
\sum_jp_j\sqrt{q_j}
\Big)^2
\leq
\Big(
\sum_jp_j^2
\Big)
\Big(
\sum_jq_j
\Big)
=
\sum_jp_j^2 .
\nonumber
\end{equation}
The bound is attained for $q_j=\frac{p_j^2}{\sum_kp_k^2}.$ Therefore,
\begin{equation}
\CA(|\psi\rangle)
=
1-\sum_jp_j^2
=
2\sum_{j<k}p_jp_k .
\label{eq:CA-formula}
\end{equation}
By contrast, Proposition~\ref{prop:pure-formula} gives
\begin{align}
\MA(|\psi\rangle)
&=
\frac12
\Big[
1-
\Big|
\sum_jp_je^{2i\theta_j}
\Big|^2
\Big]
\nonumber\\
&=
2\sum_{j<k}p_jp_k
\sin^2(\theta_j-\theta_k).
\label{eq:MA-phase-form}
\end{align}
Thus $\CA$ depends only on the population distribution in the reference basis, whereas $\MA$ also depends on the relative phases. Since $\calI\subset\calF$, the minimization for imaginarity is taken over a larger free set than the corresponding minimization for coherence. Hence
\begin{equation}
0
\leq
\MA(|\psi\rangle)
\leq
\CA(|\psi\rangle).
\label{eq:coherence-hierarchy}
\end{equation}
Equivalently, subtracting Eq.~\eqref{eq:MA-phase-form} from Eq.~\eqref{eq:CA-formula} gives
\begin{equation}
\CA(|\psi\rangle)-\MA(|\psi\rangle)
=
2\sum_{j<k}p_jp_k
\cos^2(\theta_j-\theta_k).
\label{eq:coherence-gap}
\end{equation}

The expressions for $\MA$ and $\CA$ contain the same pairwise weights $2p_jp_k$, with an additional factor $\sin^2(\theta_j-\theta_k)$ in $\MA$. Equation~\eqref{eq:coherence-gap} concerns these two pure-state functionals. It neither gives a general additive decomposition of coherence nor establishes the difference as a resource monotone. The lower equality in Eq.~\eqref{eq:coherence-hierarchy} holds exactly for states that are real up to a global phase. Nontrivial saturation of the upper equality requires all nonzero pairwise phase differences on the support to be $\pi/2$ modulo $\pi$, which can occur only for support size at most two.

\subsection{Single-qubit convex roof and standard measures}
\label{sec:qubit-comparison}
The pure state hierarchy derived above holds in arbitrary finite dimension. For mixed states, the convex roof involves optimization over pure-state ensembles. We now solve this problem for a qubit, whose non-real part in the computational basis is specified by one affine Bloch coordinate. Write
\begin{equation}
\rho
=
\frac12
\Big(
I+x\sigma_x+y\sigma_y+z\sigma_z
\Big),
\qquad
x^2+y^2+z^2\leq1 .
\nonumber
\end{equation}
In the computational basis, $\rho_{01}=\frac{1}{2}(x-iy).$ Hence a qubit state is real if and only if $y=0$. Geometrically, the free states form the $xz$ plane inside the Bloch ball.

\begin{theorem}
\label{thm:qubit-formula}
For every single-qubit state $\rho$ with Bloch vector $(x,y,z)$,
\begin{equation}
\MAcr(\rho)
=
\frac{y^2}{2}.
\label{eq:qubit-formula}
\end{equation}
Equivalently, $\MAcr(\rho)=2[\operatorname{Im}(\rho_{01})]^2 .$
\end{theorem}

{\sf [Proof]} For a pure qubit state, the Bloch vector satisfies $	x^2+y^2+z^2=1$. A direct calculation in the computational basis gives $	|\langle\psi^*|\psi\rangle|^2=1-y^2 .$ Proposition~\ref{prop:pure-formula} therefore yields
\begin{equation}
\MA(|\psi\rangle)=\frac{y^2}{2}.
\label{eq:pure-qubit-y}
\end{equation}
Let $\rho=\sum_kp_k|\psi_k\rangle\langle\psi_k|$ be an arbitrary pure state decomposition of $\rho$, and let $y_k$ denote the $y$ component of the Bloch vector of $|\psi_k\rangle$. Since the Bloch vector depends affinely on the density matrix, $y=\sum_kp_ky_k$. Using Eq.~\eqref{eq:pure-qubit-y} for each pure component and applying Jensen's inequality, we obtain
\begin{align}
\sum_kp_k\MA(|\psi_k\rangle)
&=
\frac12\sum_kp_ky_k^2
\nonumber\\
&\geq
\frac12
\left(
\sum_kp_ky_k
\right)^2
=\frac{y^2}{2}.
\nonumber
\end{align}
Because the decomposition was arbitrary, the convex roof satisfies
\begin{equation}
\MAcr(\rho)\geq\frac{y^2}{2}.
\label{eq:qubit-lower-bound}
\end{equation}
It remains to construct a decomposition that attains this lower bound. Set $R=\sqrt{1-y^2}.$ The positivity condition $x^2+y^2+z^2\leq1$ implies $x^2+z^2\leq R^2 .$ If $R=0$, then $y^2=1$ and positivity forces $x=z=0$. The state is then pure, and Eq.~\eqref{eq:pure-qubit-y} gives Eq.~\eqref{eq:qubit-formula}. Assume now that $R>0$. The point $(x,z)$ lies in the closed disk of radius $R$. Hence it can be expressed as a convex combination of boundary points $(X_k,Z_k)$ satisfying $X_k^2+Z_k^2=R^2 .$ For each such boundary point, the Bloch vector $(X_k,y,Z_k)$ satisfies $X_k^2+y^2+Z_k^2=1 .$ It therefore represents a pure qubit state. The convex combination of these pure states has Bloch vector $(x,y,z)$ and hence gives a pure state decomposition of $\rho$. All pure components in this decomposition have the same $y$ coordinate. By Eq.~\eqref{eq:pure-qubit-y}, each component has affinity-square imaginarity $\frac{1}{2}y^2$, so the average imaginarity of this decomposition is exactly $\frac{1}{2}y^2$. This proves
\begin{equation}
\MAcr(\rho)\leq\frac{y^2}{2}.
\nonumber
\end{equation}
Together with Eq.~\eqref{eq:qubit-lower-bound}, we obtain
\begin{equation}
\MAcr(\rho)=\frac{y^2}{2}.
\nonumber
\end{equation}
Using $\rho_{01}=\frac{1}{2}(x-iy)$ gives $y=-2\operatorname{Im}(\rho_{01}),$ and therefore
\begin{equation}
\MAcr(\rho)
=
2\left[\operatorname{Im}(\rho_{01})\right]^2 .
\nonumber
\end{equation}
This proves Eq.~\eqref{eq:qubit-formula}. $\hfill\square$

The proof of Eq.~\eqref{eq:qubit-formula} identifies the geometric reason for the exceptional simplicity of the qubit case. At fixed $y$, the real Bloch coordinates $(x,z)$ can vary on a circle of pure states without changing the pure state value of $\MA$. Hence the convex roof is saturated by decompositions whose pure components have the same imaginary Bloch coordinate. For qubits, the non-real part of a state is therefore encoded in a single affine parameter.

Consider the one-parameter family
\begin{equation*}
\rho(p,\phi)
=
\frac12
\begin{pmatrix}
	1 & pe^{i\phi}\\
	pe^{-i\phi} & 1
\end{pmatrix},
\qquad
0\leq p\leq1,
\end{equation*}
Its Bloch vector is $(x,y,z)=(p\cos\phi,-p\sin\phi,0)$. Equation~\eqref{eq:qubit-formula} gives
\begin{equation*}
\MAcr[\rho(p,\phi)]
=
\frac{1}{2}p^2\sin^2\phi.
\end{equation*}
This expression separates the magnitude of the off-diagonal element, controlled by $p$, from its non-real phase contribution, controlled by $\sin\phi$. When $\phi=0$ or $\pi$, the state can have off-diagonal coherence but has no imaginarity. When $p=1$ and $\phi=\pi/2$ modulo $\pi$, the off-diagonal element is purely imaginary and the maximal qubit value $\MAcr=1/2$ is attained.

The exact qubit formula also permits a direct comparison with standard imaginarity quantifiers. Existing measures include robustness, $l_1$-norm, trace-distance, and entropy-based constructions \cite{Wu17,Du2025PureState,Xu2024Tsallis,Guo2025Distance,Tian2025JSD}. The robustness of imaginarity can be written as
\begin{equation*}
R(\rho)=\frac12\|\rho-\rho^T\|_1 ,
\end{equation*}
where the transposition is taken in the fixed reference basis. This is the normalization used in Refs.~\cite{Wu1,Wu17}. For pure states, this gives
\begin{equation*}
R(|\psi\rangle)=\sqrt{1-|\langle\psi^*|\psi\rangle|^2}.
\end{equation*}
Together with the pure state formula
\begin{equation*}
\MA(|\psi\rangle)=\frac{1-|\langle\psi^*|\psi\rangle|^2}{2},
\end{equation*}
we obtain
\begin{equation}
\MA(|\psi\rangle)=\frac{R(|\psi\rangle)^2}{2}.
\label{eq:MA-robustness}
\end{equation}
For a single-qubit mixed state, Eq.~(\ref{eq:qubit-formula}) and $R(\rho)=\frac12\|\rho-\rho^T\|_1=|y|$ give
\begin{equation}
M_A^{\rm cr}(\rho)=\frac{R(\rho)^2}{2}.\label{eq:qubit-robustness-relation}
\end{equation}

The $l_1$-norm imaginarity is
\begin{equation}
M_{l_1}(\rho)
=
\sum_{i\neq j}
\left|
\operatorname{Im}(\rho_{ij})
\right|.
\label{eq:l1-imaginarity-def}
\end{equation}
For a qubit, $\rho_{01}=\frac{1}{2}(x-iy)$, and hence
\begin{equation*}
M_{l_1}(\rho)=|y|.
\end{equation*}
Combining this identity with Eq.~\eqref{eq:qubit-formula} yields
\begin{equation}
\MAcr(\rho)
=
\frac{M_{l_1}(\rho)^2}{2}.
\label{eq:l1-relation}
\end{equation}

The trace-distance imaginarity is defined as
\begin{equation}
M_{\mathrm{Tr}}(\rho)
=
\frac12
\min_{\sigma\in\calF}
\|\rho-\sigma\|_1 .
\label{eq:trace-distance-def}
\end{equation}
For a qubit, the optimal real state is obtained by removing the $y$ component and keeping the real Bloch coordinates fixed
\begin{equation*}
\sigma_{\rm opt}=\frac12(I+x\sigma_x+z\sigma_z).
\end{equation*}
Thus
\begin{equation*}
\rho-\sigma_{\rm opt}
=
\frac{y}{2}\sigma_y .
\end{equation*}
The eigenvalues of $(\frac{y}{2})\sigma_y$ are $\pm \frac{1}{2}|y|$, so
\begin{equation}
M_{\mathrm{Tr}}(\rho)
=
\frac{|y|}{2}.
\label{eq:trace-qubit}
\end{equation}
Equation~\eqref{eq:qubit-formula} then gives
\begin{equation}
\MAcr(\rho)
=
2M_{\mathrm{Tr}}(\rho)^2 .
\label{eq:trace-relation}
\end{equation}

Combining Eqs.~\eqref{eq:qubit-robustness-relation}, \eqref{eq:l1-relation}, and \eqref{eq:trace-relation}, we obtain the single-qubit identities
\begin{equation}
\MAcr(\rho)
=
\frac{R(\rho)^2}{2}
=
\frac{M_{l_1}(\rho)^2}{2}
=
2M_{\mathrm{Tr}}(\rho)^2 .
\label{eq:qubit-quadratic-identities}
\end{equation}
\begin{table*}[t]
\caption{Comparison of imaginarity measures in a fixed reference basis.}
\label{tab:imaginarity-comparison}
\centering
\small
\setlength{\tabcolsep}{5pt}
\renewcommand{\arraystretch}{1.08}
\begin{tabular}{@{}p{0.25\textwidth}p{0.15\textwidth}p{0.22\textwidth}p{0.32\textwidth}@{}}
\toprule
\raggedright Measure & \raggedright Pure state & \raggedright Mixed state & \raggedright Relation to $\MA$ or $\MAcr$ \tabularnewline
\midrule
\raggedright $\MAcr$ (this work) & \raggedright $(1-s^2)/2$ & \raggedright Qubit: $y^2/2$;\newline convex roof in general & \raggedright Optimal qubit ensemble:\newline common $y$ \tabularnewline
\addlinespace[3pt]
\raggedright Geometric $I_g$~\cite{Wu17,Kondra2023RealOps} & \raggedright $(1-s)/2$ & \raggedright $[1-q(\rho)]/2$ & \raggedright $\MA=2I_g(1-I_g)$ (pure) \tabularnewline
\addlinespace[3pt]
\raggedright Robustness $R$~\cite{Wu1,Wu17} & \raggedright $\sqrt{1-s^2}$ & \raggedright $\|\rho-\rho^*\|_1/2$ & \raggedright $\MAcr=R^2/2$ (pure; mixed qubits) \tabularnewline
\addlinespace[3pt]
\raggedright Trace distance $M_{\mathrm{Tr}}$~\cite{Hickey1,Wu17} & \raggedright $\sqrt{1-s^2}/2$ & \raggedright $\|\rho-\rho^*\|_1/4$ & \raggedright $\MAcr=2M_{\mathrm{Tr}}^2$ (pure; mixed qubits) \tabularnewline
\addlinespace[3pt]
\raggedright Tsallis-based $T_{1/2}$~\cite{Xu2024Tsallis} & \raggedright $1-s^2$ & \raggedright $1-A(\rho,\rho^*)$ & \raggedright $T_{1/2}=2\MA$ (pure) \tabularnewline
\addlinespace[3pt]
\raggedright Relative entropy $M_{\mathrm{rel}}$~\cite{Xue} & \raggedright $h_2((1+s)/2)$ & \raggedright $S(\rho_{\mathbb R})-S(\rho)$ & \raggedright Not fixed by $\MAcr$ (mixed qubits) \tabularnewline
\bottomrule
\end{tabular}
\par\smallskip
\parbox{\textwidth}{\footnotesize\raggedright Here $s=|\langle\psi^*|\psi\rangle|$, $\rho_{\mathbb R}=(\rho+\rho^*)/2$, and $q(\rho)$ is the root fidelity between $\rho$ and $\rho^*$. Pure-state formulas and relations hold in arbitrary dimension. All six measures are faithful, convex, and monotone under real operations, including on average under selective real operations. The trace-distance convention is $M_{\mathrm{Tr}}=\frac12\min_{\sigma\in\calF}\|\rho-\sigma\|_1$, with optimizer $\rho_{\mathbb R}$; the entrywise $l_1$ relation in Eq.~\eqref{eq:qubit-quadratic-identities} applies to qubits only.}
\end{table*}

Equation~\eqref{eq:MA-robustness} shows that $\MA$ orders pure states in the same way as robustness. For mixed qubits, Eq.~\eqref{eq:qubit-quadratic-identities} gives the same ordering as robustness, $l_1$-norm imaginarity, and trace-distance imaginarity. These relations do not establish a separate operational advantage. The affinity-square optimization is the $\alpha=1/2$ case of Theorem~1(i) in Ref.~\cite{Guo2025Distance}; here it supplies the pure-state input to the convex roof. Its pure-state value is half the Tsallis-based quantity at parameter $1/2$ in Ref.~\cite{Xu2024Tsallis}. Our analysis evaluates this optimization over the full real-state set, derives the superposition formulas, and gives a direct Bloch-space evaluation of the qubit convex roof with an explicit optimal ensemble. The pure-state function $f_A(s)=(1-s^2)/2$, where $s=|\langle\psi^*|\psi\rangle|$, is covered by the construction in Ref.~\cite{Du2025PureState}; the proof in Sec.~\ref{sec:affinity} verifies its monotonicity directly. The qubit value also follows from Theorems~5 and 6 of that reference, since their expression for this choice of $f_A$ is convex.

The relative entropy of imaginarity has a different dependence. It is
\begin{equation*}
M_{\mathrm{rel}}(\rho)
=
S(\rho_{\mathbb R})-S(\rho),
\qquad
\rho_{\mathbb R}
=
\frac{\rho+\rho^*}{2},
\label{eq:relative-entropy-imaginarity}
\end{equation*}
where $S(\rho)=-\Tr(\rho\log\rho)$. For a pure state,
\begin{equation}
M_{\mathrm{rel}}(|\psi\rangle)
=
h_2
\Big(
\frac{1+|\langle\psi^*|\psi\rangle|}{2}
\Big),
\label{eq:relative-pure}
\end{equation}
with $h_2(u)=-u\log u-(1-u)\log(1-u).$ All logarithms are taken in base 2. Using the pure state formula for $\MA$, Eq.~\eqref{eq:relative-pure} becomes
\begin{equation*}
M_{\mathrm{rel}}(|\psi\rangle)
=
h_2
\Big(
\frac{
	1+\sqrt{1-2\MA(|\psi\rangle)}
}{2}
\Big).
\label{eq:relative-pure-MA}
\end{equation*}

For a single-qubit mixed state, define
\begin{equation*}
r=\sqrt{x^2+y^2+z^2},
\qquad
r_{\mathbb R}=\sqrt{x^2+z^2}.
\end{equation*}
The eigenvalues of $\rho$ and $\rho_{\mathbb R}$ are $\frac{1}{2}(1\pm r)$ and $\frac{1}{2}(1\pm r_{\mathbb R})$, respectively. Hence
\begin{equation*}
M_{\mathrm{rel}}(\rho)=h_2\Big(\frac{1+r_{\mathbb R}}{2}\Big)-h_2\Big(\frac{1+r}{2}\Big).
\label{eq:relative-qubit}
\end{equation*}
Unlike robustness, $l_1$-norm imaginarity, and trace-distance imaginarity, $M_{\mathrm{rel}}$ depends on both the non-real coordinate $|y|$ and the real Bloch components through $r_{\mathbb R}$. Therefore, for mixed qubit states, the relative entropy of imaginarity is not determined by $\MAcr(\rho)$ alone.

Table~\ref{tab:imaginarity-comparison} compares the pure- and mixed-state expressions of several imaginarity measures. We write $s=|\langle\psi^*|\psi\rangle|$ and use $q(\rho)=\Tr\sqrt{\sqrt{\rho}\rho^*\sqrt{\rho}}$ for the root fidelity between $\rho$ and $\rho^*$. Another related choice is the geometric-like pure-state function $1-\sqrt{(1+s)/2}$~\cite{Guo2024}. The Tsallis-based quantity at parameter $1/2$ is $T_{1/2}(\rho)=1-A(\rho,\rho^*)$~\cite{Xu2024Tsallis}. It equals $2\MA$ on pure states but is defined directly on mixed states, without an ensemble optimization.

\section{Summary and discussion}
We have studied a convex-roof affinity-square measure of quantum imaginarity with respect to a fixed reference basis. The pure state quantity is defined by minimizing the affinity-square measure over the full set of real density matrices, and this optimization admits a closed expression in terms of the conjugation overlap between a state and its complex conjugate. Using the pure state expression, we proved that the convex-roof extension is faithful, convex, monotone under real operations, and nonincreasing on average under selective real operations. These properties establish $\MAcr$ as a valid monotone in the resource theory of imaginarity. The proof also shows why the squared-affinity form is useful in this setting: it converts the pure state optimization over the real-state set into an explicit phase-sensitive expression, which then provides a tractable input for the convex-roof construction.

The pure-state relation between $\MA$ and $\CA$ follows from the factors $\sin^2(\theta_j-\theta_k)$ in their pairwise expansions. For arbitrary single-qubit mixed states, the convex roof reduces to a closed expression determined by the imaginary part of the off-diagonal density-matrix element, or equivalently by the displacement of the Bloch vector from the real-state plane. This result gives a geometric interpretation of qubit imaginarity and yields explicit quadratic relations with robustness, the $l_1$-norm imaginarity, and trace-distance imaginarity. The relative entropy of imaginarity behaves differently, since it also depends on the real Bloch components and is therefore not determined by $\MAcr$ alone for mixed qubit states. We obtain a closed mixed-state expression for qubits; a general formula in higher dimensions is not derived here. These results provide an affinity-based characterization of imaginarity and identify the role of conjugation-induced phase structure in non-real quantum resources.

\begin{acknowledgments}
This work was supported by the National Natural Science Foundation of China under Grant No.~12301582; the Jiangxi Provincial Natural Science Foundation under Grant No.~20252BAC200156; the Natural Science Foundation of Shandong Province under Grant No.~ZR2024MA068; GDSTA:SKXRC2025442; the Early-Career Young Scientists and Technologists Project of Jiangxi Province under Grant No.~20244BCE52197; the Science and Technology Project of the Jiangxi Provincial Department of Education under Grant No.~GJJ2400606; and the Doctoral Research Startup Fund Project of East China University of Technology under Grant No.~DHBK2024029.
\end{acknowledgments}

\end{document}